  \documentstyle[preprint,aps]{revtex} 
  \tightenlines                                    
\newcommand{\bra}[1]{\langle#1|}
\newcommand{\ket}[1]{|#1\rangle}

\begin{document}
\draft
\preprint{}
\title{Operational Characterization of Simultaneous Measurements
in Quantum Mechanics}
\author{Masanao Ozawa}
\address{School of Informatics and Sciences,
Nagoya University, Chikusa-ku, Nagoya 464-01, Japan}
\date{\today}
\maketitle
\begin{abstract}
Quantum mechanics predicts the joint probability distribution 
of the outcomes of simultaneous measurements of commuting observables,
but, in the state of the art, has lacked the operational definition of
simultaneous measurements.
The question is answered as to when the consecutive applications of
measuring apparatuses give a simultaneous measurement of their
observables.
For this purpose, all the possible state reductions caused by
measurements of an observable is also characterized by their
operations. 
\end{abstract}
\narrowtext
In quantum mechanics, observables are represented by linear operators, 
for which the product operation is not commutative.
If two observables are represented by commuting operators,
they are simultaneously measurable and quantum mechanics
predicts the joint probability distribution of the outcomes
of their simultaneous measurement.
But, it has not been answered fully what measurement can be considered
as a simultaneous measurement of those observables. 

Let ${\bf S}$ be a quantum system with the Hilbert space ${\cal H}$ of
state vectors.
Let $A$ be an observable of ${\bf S}$. 
For any Borel set $\Delta$ in the real line ${\rm R}$, 
the spectral projection of $A$ corresponding 
to $\Delta$ is denoted by $E^{A}(\Delta)$; if $A$ has the Dirac type
spectral representation
\begin{equation}
A=\sum_{\nu}\sum_{\mu}\mu\ket{\mu,\nu}\bra{\mu,\nu}
+\sum_{\nu}\int\lambda\ket{\lambda,\nu}\bra{\lambda,\nu}\,d\lambda,
\end{equation}
where $\mu$ varies over the discrete eigenvalues, $\lambda$ varies over
the continuous eigenvalues, and $\nu$ is the degeneracy parameter,
then we have
\begin{equation}
E^{A}(\Delta)=\sum_{\nu}\sum_{\mu\in\Delta}\ket{\mu,\nu}\bra{\mu,\nu}
+\sum_{\nu}\int_{\Delta}\ket{\lambda,\nu}\bra{\lambda,\nu}\,d\lambda.
\end{equation}
For any real number $t$, we shall denote by $\rho(t)$ the 
state (density operator) of ${\bf S}$ at the time $t$. 
We shall denote by ``$A(t)\in\Delta$'' the probabilistic event that
{\em the outcome of the measurement (under consideration) of 
an observable $A$ at the time $t$ is in a Borel set $\Delta$}.
According to the Born statistical formula, we have
\begin{equation}\label{eq:B1}
\Pr\{A(t)\in\Delta\}={\rm Tr}[E^{A}(\Delta)\rho(t)].
\end{equation}

Any commuting observables $A$ and $B$ are simultaneously measurable 
and the joint probability distribution of the outcomes of their 
simultaneous measurement is postulated by
\begin{equation}\label{eq:13}
\Pr\{A(t)\in\Delta,B(t)\in\Delta'\}
={\rm Tr}[E^{A}(\Delta)E^{B}(\Delta')\rho(t)].
\end{equation}

A well-known proof of this quantum rule runs as follows\cite{vN55}. 
Since $A$ and $B$ commute,
there exist an observable $C$ and Borel functions
$f$ and $g$ such that $A=f(C)$ and $B=g(C)$. 
For the outcome $c$ of the $C$-measurement, 
define the outcome of the $A$-measurement to be $f(c)$ and 
the outcome of the $B$-measurement to be $g(c)$.
Then, 
it follows from (\ref{eq:B1}) that their outcomes satisfy (\ref{eq:13})
 so that
the measurement of $C$ at the time $t$ gives 
a simultaneous measurement of $A$ and $B$.

The above proof gives one special instance of simultaneous measurement
which uses only one measuring apparatus, 
but it is rather open when a pair of measuring apparatuses for 
$A$ and $B$ make a simultaneous measurement of $A$ and $B$.
The purpose of this paper is to answer this question.

The conventional approach to the problem assumes the {\em projection 
postulate} formulated by L\"{u}ders \cite{Lud51} as follows:
{\em If an observable $A$ is measured in a state $\rho$, then
at the time just after measurement the object is left in
the state 
$$
\frac{E^{A}(\{a\})\rho E^{A}(\{a\})}{{\rm Tr}[E^{A}(\{a\})\rho]},
$$
provided that the object leads to the outcome $a$.}

Suppose that at time $t$ an observable $A$ is measured by an 
apparatus satisfying the projection postulate and that 
at the time, $t+\Delta t$, just after the $A$-measurement an arbitrary
observable $B$ is measured.
The joint probability distribution of the outcomes of
the $A$-measurement and the $B$-measurement is given by
\begin{eqnarray}
\lefteqn{\Pr\{A(t)\in\Delta,B(t+\Delta t)\in\Delta'\}}\qquad\nonumber\\
&=&\sum_{a\in\Delta}
{\rm Tr}[E^{B}(\Delta)E^{A}(\{a\})\rho(t)E^{A}(\{a\})].\label{eq:804c}
\end{eqnarray}
Thus, if $A$ and $B$ commute, we have
\begin{equation}
\Pr\{A(t)\in\Delta,B(t+\Delta t)\in\Delta'\}
={\rm Tr}[E^{A}(\Delta)E^{B}(\Delta')\rho(t)].\label{eq:804d}
\end{equation}
We can therefore reinterpret the $B$-measurement at the time $t+\Delta t$
as the $B$-measurement at the time $t$ and consider the above
consecutive measurements of $A$ and $B$ as the simultaneous measurement
of $A$ and $B$ at the time $t$.

If we would restrict measurements to those satisfying the projection 
postulate, 
any consecutive measurements of commuting observables could
be considered as simultaneous measurement,
but this approach has the following limitations:

(i) Some of the most familiar measuring apparatuses such as photon 
counters do not satisfy the projection postulate \cite{IUO90}.

(ii) When the observable has continuous spectrum, no measurements
satisfy the projection postulate \cite{84QC}.

The above limitations appear indeed quite serious.
From (i), we cannot apply
(\ref{eq:13}) to correlation measurements using photon counters as in
most of optical experiments and EPR-correlation measurements
\cite{MW95,AGR82,ADR82,BPMEWZ97}.
From (ii), we cannot use
(\ref{eq:13}) for any measurements of continuous observables using more
than one apparatuses.

Now we shall abandon the projection postulate and 
consider the following problem:
{\em under what condition can consecutive measurements
of two or more observables be considered as a simultaneous
measurement of those observables?}

Suppose that an observable $A$ of ${\bf S}$ is measured at time $t$
by a measuring apparatus ${\bf A}$ with the state space ${\cal H}_{{\bf A}}$.
Let $t+\Delta t$ be the time just after measurement.
The measurement is carried out by the interaction between ${\bf S}$ 
and ${\bf A}$ from $t$ to $t+\Delta t$,
and after the time $t+\Delta t$ the object ${\bf S}$ is free from the
apparatus ${\bf A}$.
Suppose that the apparatus is in the state $\sigma$ at the time $t$
and let $U$ be the unitary operator representing the 
time evolution of the composite system ${\bf S}+{\bf A}$ from $t$ 
to $t+\Delta t$.
Then, the object ${\bf S}$ is in the state
\begin{equation}\label{eq:803a}
\rho(t+\Delta t)={\rm Tr}_{{\bf A}}[U(\rho(t)\otimes\sigma)U^{\dagger}]
\end{equation}
at the time $t+\Delta t$, where ${\rm Tr}_{{\bf A}}$ is the partial trace
over ${\cal H}_{{\bf A}}$.
The state change $\rho(t)\mapsto\rho(t+\Delta t)$ is determined 
independent of the outcome of measurement and called
the {\em nonselective state change}.

Let $B$ be an arbitrary observable of ${\bf S}$.
We say that the measurement of $A$ using the apparatus ${\bf A}$
{\em does not disturb} the observable $B$ iff the nonselective
state change does not perturb the probability distribution of $B$,
that is, we have
\begin{equation}\label{eq:1224a}
{\rm Tr}[E^{B}(\Delta)e^{-iH\Delta t/\hbar}\rho(t)e^{iH\Delta t/\hbar}]
={\rm Tr}[E^{B}(\Delta)\rho(t+\Delta t)]
\end{equation}
for any Borel set $\Delta$, where $H$ is the Hamiltonian 
of the system ${\bf S}$.
The measurement is said to be {\em instantaneous} iff the duration 
$\Delta t$ of measuring interaction is negligible in the time scale 
of the time evolution of the object.
Thus, the instantaneous measurement of $A$ using ${\bf A}$ does not
disturb $B$ if and only if
\begin{equation}\label{eq:10}
{\rm Tr}[E^{B}(\Delta)\rho(t)]={\rm Tr}[E^{B}(\Delta)\rho(t+\Delta t)]
\end{equation}
for any Borel set $\Delta$.

By (\ref{eq:803a}) and by the property of the partial trace, we have
\begin{eqnarray}
\lefteqn{{\rm Tr}[E^{B}(\Delta)\rho(t+\Delta t)]}\qquad\nonumber\\
&=&{\rm Tr}\left[E^{B}(\Delta)
{\rm Tr}_{{\bf A}}[U(\rho(t)\otimes\sigma)U^{\dagger}]\right]\nonumber\\
&=&{\rm Tr}\left[
{\rm Tr}_{{\bf A}}[U^{\dagger}(E^{B}(\Delta)\otimes I)U(I\otimes\sigma)]
\rho(t)\right].
\label{eq:108e}
\end{eqnarray}
Since $\rho(t)$ is arbitrary, (\ref{eq:10}) is equivalent to
\begin{equation}\label{eq:11}
E^{B}(\Delta)
={\rm Tr}_{{\bf A}}[U^{\dagger}(E^{B}(\Delta)\otimes I)U(I\otimes\sigma)]
\end{equation}
for any Borel set $\Delta$.

Now we can state precisely the answer to the above problem to be obtained 
in the present paper.

{\bf Theorem 1.}
{\em The instantaneous measurement of an observable $A$ at time $t$ 
using an apparatus ${\bf A}$ does not disturb an observable $B$ 
if and only if we have
\begin{eqnarray}\label{eq:12}
\lefteqn{\Pr\{A(t)\in\Delta,B(t+\Delta t)\in\Delta'\}}\qquad\nonumber\\
&=&{\rm Tr}[E^{A}(\Delta)E^{B}(\Delta')\rho(t)]
\end{eqnarray}
for any density operator $\rho(t)$ and any Borel sets
$\Delta$ and $\Delta'$.
In this case, $A$ and $B$ necessarily commute.}

In what follows, 
we shall characterize all the possible state reductions caused
by measurements of an observables 
in order to provide the proof of the above theorem, 

For any Borel set $\Delta$, 
let $\rho(t+\Delta t|A(t)\in\Delta)$ be the state at $t+\Delta t$ 
of the object ${\bf S}$ 
conditional upon $A(t)\in\Delta$.
Thus, if the object ${\bf S}$ is sampled randomly from the subensemble 
of the similar systems that yield the outcome of the 
$A$-measurement in the Borel set $\Delta$, then ${\bf S}$ is 
in the state $\rho(t+\Delta t|A(t)\in\Delta)$ at the time $t+\Delta t$.
When $\Delta={\rm R}$, the condition $A(t)\in\Delta$ makes no selection,
and hence we have
\begin{equation}\label{eq:108b}
\rho(t+\Delta t|A(t)\in{\rm R})=\rho(t+\Delta t).
\end{equation}
For $\Delta\not={\rm R}$,
the state change $\rho(t)\mapsto\rho(t+\Delta t|A(t)\in\Delta)$ is called
the {\em selective state change}.
When $\Pr\{A(t)\in\Delta\}=0$, the state $\rho(t+\Delta t|A(t)\in\Delta)$ is
indefinite, and let $\rho(t+\Delta t|A(t)\in\Delta)$ be an arbitrarily chosen 
density operator for mathematical convenience.

Davies and Lewis \cite{DL70} postulated that for any Borel set $\Delta$ 
there exists a positive linear transformation $T_{\Delta}$ on the space
${\tau c}({\cal H})$ of trace class operators on ${\cal H}$
satisfying the following conditions:

(i) For any Borel set $\Delta$ and disjoint Borel sets $\Delta_{n}$ such that
$\Delta=\bigcup_{n}\Delta_{n}$ and for any $\rho\in{\tau c}({\cal H})$, 
\begin{mathletters}
\begin{equation}\label{eq:1230a}
T_{\Delta}(\rho)=\sum_{n}T_{\Delta_{n}}(\rho).
\end{equation}

(ii) For any Borel set $\Delta$ and any $\rho\in{\tau c}({\cal H})$,
\begin{equation}\label{eq:1230c}
{\rm Tr}[T_{\Delta}(\rho)]={\rm Tr}[E^{A}(\Delta)\rho].
\end{equation}

(iii) For any Borel set $\Delta$ with $\Pr\{A(t)\in\Delta\}>0$,
\begin{equation}\label{eq:1230d}
\rho(t+\Delta t|A(t)\in\Delta)
=\frac{T_{\Delta}[\rho(t)]}{{\rm Tr}[T_{\Delta}[\rho(t)]]}.
\end{equation}
\end{mathletters}

The validity of the Davies-Lewis postulate was previously 
demonstrated in \cite{84QC} {\em based on} the joint probability formula,
where it is also shown that any such transformations $T_{\Delta}$ which are 
realizable by a measuring process are completely positive and 
{\em vice versa}.
In what follows, we shall prove the Davies-Lewis 
postulate {\em without} assuming the joint probability formula so as
to avoid the circular argument.

Suppose that at the time $t+\Delta t$ the observer were to  
measure an arbitrary observable $B$ of the same object ${\bf S}$.
Then, the joint probability distribution of the outcomes
of the $A$-measurement and the $B$-measurement satisfies
\begin{eqnarray}\label{eq:8}
\lefteqn{\Pr\{A(t)\in\Delta,B(t+\Delta t)\in\Delta'\}}\quad\nonumber\\
&=&{\rm Tr}[E^{B}(\Delta')\rho(t+\Delta t|A(t)\in\Delta)]
{\rm Tr}[E^{A}(\Delta)\rho(t)].
\end{eqnarray}
For any Borel set $\Delta$, 
let $T_{\Delta}[\rho(t)]$ be the trace class operator defined by
\begin{equation}\label{eq:T2}
T_{\Delta}[\rho(t)]
={\rm Tr}[E^{A}(\Delta)\rho(t)]\rho(t+\Delta t|A(t)\in\Delta).
\end{equation}
Then, (\ref{eq:T2}) defines the transformation $T_{\Delta}$ that
maps $\rho(t)$ to $T_{\Delta}[\rho(t)]$.
It follows from (\ref{eq:8}) and (\ref{eq:T2}) that $T_{\Delta}$ satisfies
\begin{equation}\label{eq:T3}
\Pr\{A(t)\in\Delta,B(t+\Delta t)\in\Delta'\}
={\rm Tr}[E^{B}(\Delta')T_{\Delta}[\rho(t)]].
\end{equation}
Suppose that the state $\rho(t)$ is a mixture of 
the states $\rho_{1}$ and
$\rho_{2}$, i.e., 
\begin{equation}\label{eq:T4}
\rho(t)=\alpha\rho_{1}+(1-\alpha)\rho_{2}
\end{equation}
where $0<\alpha<1$.
This means that at the time $t$ 
the measured object ${\bf S}$ is sampled 
randomly from an ensemble of similar systems described by the
density operator $\rho_{1}$ with probability $\alpha$ and from 
another ensemble described by the density operator $\rho_{2}$
with probability $1-\alpha$.
Thus we have naturally
\begin{eqnarray}\label{eq:T4.5}
\lefteqn{\Pr\{A(t)\in\Delta,B(t+\Delta t)\in\Delta'|\rho(t)
=\alpha\rho_{1}+(1-\alpha)\rho_{2}\}}\nonumber\\
&=&\alpha\Pr\{A(t)\in\Delta,B(t+\Delta t)\in\Delta'|\rho(t)
=\rho_{1}\}\nonumber\\
& &\mbox{ }+(1-\alpha)\Pr\{A(t)\in\Delta,B(t+\Delta t)\in\Delta'|\rho(t)
=\rho_{2}\},\nonumber\\
\end{eqnarray}
where $\Pr\{E|F\}$ stands for the conditional probability
of $E$ given $F$.   
From (\ref{eq:T3}) and (\ref{eq:T4.5}), we have
\begin{eqnarray}\label{eq:T5}
\lefteqn{{\rm Tr}\left[E^{B}(\Delta')T_{\Delta}
\left[\alpha\rho_{1}+(1-\alpha)\rho_{2}\right]\right]}\quad\nonumber\\
&=&{\rm Tr}
\left[E^{B}(\Delta')
  \left[\alpha T_{\Delta}(\rho_{1})+(1-\alpha)T_{\Delta}(\rho_{2})
  \right]
\right].
\end{eqnarray}
Since $B$ and $\Delta'$ are arbitrary, we have
\begin{equation}\label{eq:T6}
T_{\Delta}\left[\alpha\rho_{1}+(1-\alpha)\rho_{2}\right]
=\alpha T_{\Delta}(\rho_{1})+(1-\alpha)T_{\Delta}(\rho_{2}).
\end{equation}
Thus, $T_{\Delta}$ is an affine transformation from the space of density 
operators to the space of trace class operators, 
and hence it can be extended to a unique positive 
linear transformation of the trace class operators \cite{Kad65}.

We have proved that for any apparatus ${\bf A}$ measuring $A$
there is uniquely a family $\{T_{\Delta}|\ \Delta\in{\cal B}({\rm R})\}$
of positive linear transformations of the trace class operators 
such that (\ref{eq:T2}) and (\ref{eq:T3}) hold,
where ${\cal B}({\rm R})$ stands for the collection of all Borel sets.
This family of linear transformations will be referred to
as the {\em operational distribution} of the apparatus ${\bf A}$.

By the countable additivity of probability, 
if $\Delta=\bigcup_{n}\Delta_{n}$ for disjoint Borel sets $\Delta_{n}$, 
we have
\begin{eqnarray}
\lefteqn{
\Pr\{A(t)\in\Delta,B(t+\Delta t)\in\Delta'\}
}\qquad\nonumber\\
&=&
\sum_{n}\Pr\{A(t)\in\Delta_{n},B(t+\Delta t)\in\Delta'\}.
\label{eq:108a}
\end{eqnarray}
By (\ref{eq:T3}) and (\ref{eq:108a}) we can prove (\ref{eq:1230a}).
Equations (\ref{eq:1230c}) and (\ref{eq:1230d}) 
are obvious from (\ref{eq:T2}).
Thus, the operational distribution 
$\{T_{\Delta}|\ \Delta\in{\cal B}({\rm R})\}$ 
satisfies the Davies-Lewis postulate.

\sloppy
Mathematical theory of operational distributions was introduced
by Davies and Lewis \cite{DL70}
based on relation (\ref{eq:1230a}) as a mathematical axiom
and developed in \cite{Dav76}.
Theory of measuring processes based on completely positive
operational distributions was developed extensively in 
\cite{84QC,85CA,85CC,86IQ,88MS,89RS,90QP,93CA}.

Now we are ready to state the following important relations
for operational distributions.

{\bf Theorem 2.}
{\em Let $\{T_{\Delta}|\ \Delta\in{\cal B}({\rm R})\}$ 
be a family of positive
linear transformations on ${\tau c}({\cal H})$ satisfying (\ref{eq:1230a}) 
and (\ref{eq:1230c}).
Then, for any Borel set $\Delta$ and any trace class operator
$\rho$ we have
\begin{eqnarray}\label{eq:127b}
T_{\Delta}(\rho)
&=&T_{{\rm R}}\left(E^{A}(\Delta)\rho\right)
=T_{{\rm R}}\left(\rho E^{A}(\Delta)\right)\nonumber\\
&=&T_{{\rm R}}\left(E^{A}(\Delta)\rho E^{A}(\Delta)\right).
\end{eqnarray}
}

A proof of the above theorem was given in \cite{84QC} for the 
case where $T_{\Delta}$ is completely positive, and another proof
was given in \cite{97OQ} for the case where $A$ is discrete.
The general proof necessary for the above theorem is obtained by
modifying the above proofs.

By (\ref{eq:803a}), (\ref{eq:108b}), and (\ref{eq:T2}) we have
\begin{equation}\label{eq:108c}
T_{{\rm R}}(\rho)={\rm Tr}[U(\rho\otimes\sigma)U^{\dagger}]
\end{equation}
for any $\rho\in{\tau c}({\cal H})$.
From (\ref{eq:1230d}), (\ref{eq:127b}), and (\ref{eq:108c}), we obtain
the following characterization of the possible selective state changes:
{\em if $\Pr\{A(t)\in\Delta\}>0$,
the state $\rho(t+\Delta t|A(t)\in\Delta)$ is uniquely determined as
\begin{mathletters}
\label{eq:108d}
\begin{eqnarray}
\lefteqn{\rho(t+\Delta t|A(t)\in\Delta)}\qquad\nonumber\\
&=&\frac{{\rm Tr}_{{\bf A}}[U(\rho(t)E^{A}(\Delta)\otimes\sigma)U^{\dagger}]}
{{\rm Tr}[E^{A}(\Delta)\rho(t)]}\label{eq:108d-a}\\
&=&\frac{{\rm Tr}_{{\bf A}}[U(E^{A}(\Delta)\rho(t)\otimes\sigma)U^{\dagger}]}
{{\rm Tr}[E^{A}(\Delta)\rho(t)]}\label{eq:108d-b}\\
&=&\frac{{\rm Tr}_{{\bf A}}[U(E^{A}(\Delta)\rho(t)E^{A}(\Delta)
\otimes\sigma)U^{\dagger}]}
{{\rm Tr}[E^{A}(\Delta)\rho(t)]}.\label{eq:108d-c}
\end{eqnarray}
\end{mathletters}
}

Now, we are ready to prove our main theorem.

{\em Proof of Theorem 1.}
It suffices to show the equivalence between (\ref{eq:11}) and (\ref{eq:12}).
First, note that from  (\ref{eq:8}) and (\ref{eq:108d-b}), 
the joint probability distribution of $A$ and $B$ is given by
\begin{eqnarray}
\lefteqn{\Pr\{A(t)\in\Delta,B(t+\Delta t)\in\Delta'\}}\qquad\nonumber\\
&=&
{\rm Tr}\left[E^{B}(\Delta')
{\rm Tr}_{{\bf A}}[U(E^{A}(\Delta)\rho(t)\otimes\sigma)U^{\dagger}]
\right]\nonumber\\
&=&
{\rm Tr}\left[E^{A}(\Delta)
{\rm Tr}_{{\bf A}}[U^{\dagger}(E^{B}(\Delta')\otimes I)U(I\otimes\sigma)]
\rho(t)\right].\nonumber\\
\label{eq:1016-3}
\end{eqnarray}
If (\ref{eq:11}) holds, (\ref{eq:12}) follows from (\ref{eq:1016-3}).
Conversely, suppose that (\ref{eq:12}) holds.
By substituting $\Delta={\rm R}$ in (\ref{eq:12}),
we have
\begin{equation}\label{eq:1016-4}
\Pr\{A(t)\in{\rm R},B(t+\Delta t)\in\Delta'\}={\rm Tr}[E^{B}(\Delta')\rho(t)].
\end{equation}
On the other hand, from (\ref{eq:1016-3}) we have
\begin{eqnarray}\label{eq:1016-5} 
\lefteqn{\Pr\{A(t)\in{\rm R},B(t+\Delta t)\in\Delta'\}}\qquad\nonumber\\
&=&
{\rm Tr}\left[
{\rm Tr}_{{\bf A}}[U^{\dagger}(E^{B}(\Delta)\otimes I)U(I\otimes\sigma)]
\rho(t)\right].
\end{eqnarray}
Since $\rho(t)$ is arbitrary, from (\ref{eq:1016-4}) and (\ref{eq:1016-5}) we
obtain (\ref{eq:11}).
Therefore, (\ref{eq:11}) and (\ref{eq:12}) are equivalent.
To prove $A$ and $B$ commute, suppose that (\ref{eq:12}) holds.
By the positivity of probability, the right hand side is positive.
Since $\rho(t)$ is arbitrary, the product
$E^{A}(\Delta)E^{B}(\Delta')$ is a positive self-adjoint operator so that
$E^{A}(\Delta)$ and $E^{B}(\Delta')$ commute.
Since $\Delta$ and $\Delta'$ are arbitrary, $A$ and $B$ commute.
$\Box$

Obviously from (\ref{eq:11}), if $U$ and $B\otimes I$ commute, i.e., 
\begin{equation}\label{eq:110c}
[U,E^{B}(\Delta)\otimes I]=0
\end{equation}
for any Borel set $\Delta$,
then the $A$-measurement does not disturb the observable $B$.
However, (\ref{eq:110c}) is not a necessary condition for nondisturbing
measurement.  In the case where $\sigma$
is a pure state $\sigma=\ket{\xi}\bra{\xi}$, it follows from
(\ref{eq:11}) that the $A$-measurement does not disturb the 
observable $B$ if and only if
\begin{equation}\label{eq:110e}
[U,E^{B}(\Delta)\otimes I]\ket{\psi}\ket{\xi}=0
\end{equation}
for any Borel set $\Delta$ and any state vector $\ket{\psi}$ of 
${\bf S}$.

Consider the case where the system ${\bf S}$ consists of two
subsystems ${\bf S}_{1}$ with state space ${\cal H}_{1}$ and ${\bf S}_{2}$
with state space ${\cal H}_{2}$ so that 
${\cal H}={\cal H}_{1}\otimes{\cal H}_{2}$.
Suppose that $A$ belongs to ${\bf S}_{1}$ and $B$ belongs to ${\bf S}_{2}$
so that $A=C\otimes I$ and $B=I\otimes D$ for some $C$ on ${\cal H}_{1}$
and for some $D$ on ${\cal H}_{2}$.
In this case, we say that the $A$-measurement using the apparatus ${\bf A}$
is {\em local\/} iff
the measuring interaction couples only ${\bf S}_{1}$ and ${\bf A}$.
Thus, the $A$-measurement is instantaneous and local if and
only if we have
\begin{equation}\label{eq:110d}
[U,I_{1}\otimes X\otimes I]=0,
\end{equation}
for any operator $X$ on ${\cal H}_{2}$, 
where $I_{1}$ is the identity on ${\cal H}_{1}$ and $I$ is the identity on
${\cal H}_{{\bf A}}$.
From (\ref{eq:110c}) and (\ref{eq:110d}), any local instantaneous measurement 
of $A$ does not disturb $B$ and any local instantaneous measurement
of $B$ does not disturb $A$ either.  
Therefore, it is concluded that {\em any pair of
local instantaneous measurements of $A=I\otimes C$ and 
$B=I\otimes D$ satisfies the joint probability formula 
\begin{equation}
\Pr\{A(t)\in\Delta,B(t)\in\Delta'\}
={\rm Tr}[(E^{C}(\Delta)\otimes E^{D}(\Delta'))\rho(t)],
\end{equation}
regardless of the order of the measurement.}

In the EPR paper \cite{EPR35}, the so called EPR correlation
is derived theoretically under the assumption that the pair of 
measurements satisfies the projection postulate, but the present 
result concludes that the EPR correlation holds for any pair of
local instantaneous measurements as experiments have already suggested.

\end{document}